\documentclass[conference]{IEEEtran}
\IEEEoverridecommandlockouts
\usepackage{cite}
\usepackage{amsmath,amssymb,amsfonts}
\usepackage{algorithmic}
\usepackage{graphicx}
\usepackage{textcomp}
\usepackage{xcolor}
\usepackage{subfigure}
\usepackage{multirow}
\usepackage{graphicx}

\usepackage{float}
\usepackage{comment}
\usepackage{atbegshi,picture}
\usepackage{fancyhdr}
\usepackage{eso-pic}%

\AtBeginShipoutNext{\AtBeginShipoutUpperLeft{
\put(\dimexpr\paperwidth-20.5cm\relax,-27.0cm){\begin{minipage}{555pt}{© 2022 IEEE.\small{ Personal use of this material is permitted. Permission from IEEE must be obtained for all other uses, in any current or future media, including reprinting/republishing this material for advertising or promotional purposes, creating new collective works, for resale or redistribution to servers or lists, or reuse of any copyrighted component of this work in other works.}} \end{minipage} } }  }

\def\BibTeX{{\rm B\kern-.05em{\sc i\kern-.025em b}\kern-.08em
    T\kern-.1667em\lower.7ex\hbox{E}\kern-.125emX}}
\begin{document}

\title{Explaining Hierarchical Features in Dynamic Point Cloud  Processing}

\author{\IEEEauthorblockN{ Pedro Gomes}
\IEEEauthorblockA{\textit{Dept. Electronic \& Electrical Eng.} \\
\textit{University College of London}\\
London, United Kingdom }
\and
\IEEEauthorblockN{Silvia Rossi}
\IEEEauthorblockA{\textit{Distributed Interactive System} \\
\textit{Centrum Wiskunde \& Informatica}\\
Amsterdam, The Netherlands 
}
\and
\IEEEauthorblockN{Laura Toni}
\IEEEauthorblockA{\textit{Dept. Electronic \& Electrical Eng.} \\
\textit{University College of London}\\
London, United Kingdom
}

}

\definecolor{cadmiumgreen}{rgb}{0.0, 0.42, 0.24}

\newcommand{\red}[1] {\textcolor[rgb]{1.0,0.0,0.0}{{#1}}}
\newcommand{\blue}[1] {\textcolor[rgb]{0.0,0.0,1.0}{{#1}}}
\newcommand{\cadmiumgreen}[1] {\textcolor[rgb]{0.0, 0.42, 0.24}{{#1}}}
\newcommand{\lt}[1] {\{\textbf{LT:}{\textcolor[rgb]{1,0.0,0.0}{{{\textit{{#1}}}\}}}}} 
\newcommand{\cinnamon}[1] {\textcolor[rgb]{0.82, 0.41, 0.12}{{#1}}}

\newcommand{\notePG}[1] {\{\textbf{PG:}\blue{\textit{{#1}}}\}}
\newcommand{\noteSR}[1] {\{\textbf{SR:}\cadmiumgreen{\textit{{#1}}}\}}
\newcommand{\ie}[0]{\textit{i.e.},~}
\newcommand{\eg}[0]{\textit{e.g.},~}
\newcommand{\cf}[0]{\textit{c.f.},~}

\newcommand{\etal}[0]{\textit{et al.}~}

\setlength{\intextsep}{-5pt}

\maketitle

\begin{abstract}
This paper aims at bringing some light and understanding to the field of deep learning for dynamic point cloud processing. Specifically, we focus on the hierarchical features learning aspect, with the ultimate goal of understanding which features are learned at the different stages of the process and what their meaning is. Last, we bring clarity on how hierarchical components of the network affect the learned features and their importance for a successful learning model. This study is conducted for point cloud prediction tasks, useful for predicting coding applications. 



\end{abstract}


\section{Introduction}

One major open challenge in multimedia processing is learning spatio-temporal features  for dynamic point cloud (PC) sequences. Being able to extract such information can be essential for future compression algorithms. By learning spatio-temporal features, a predictive  motion-compensated coding approach can  reduce inter-frames redundancies from the compressed bitstream~\cite{thanou2016graph}. Similarly, a PC predictor can be used as  learning-based decoder~\cite{akhtar2022inter}.
More at large, spatio-temporal features are important also in high-level PC downstream  tasks such as action recognition, prediction and obstacle avoidance.
As of today, one of the most successful methodology  is  to learn features via deep neural networks applied to each point (or group of points) instead of the whole PC. This enables the consumption of the raw PC data directly, without  pre-processing steps (\eg voxelization) that could obscure natural invariances of the data or introduce quantization errors. An example is the  pioneer  PointNet~\cite{qi2017pointnet} architecture, which learns   global PC features by aggregating local  spatial features learned by  processing each point independently.

\begin{figure}[t!]
    \centering
    \includegraphics[width=0.5\textwidth]{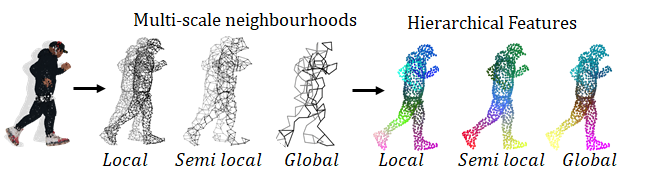}
    \caption{Hierarchical learning of features. 
    The network processes a dynamic PC at progressively larger scales to learn features.  The learned features are represented as point color using principal component analysis (PCA).}
    \label{fig:teaser}
    \vspace*{-5mm}
\end{figure}

However, in such architecture, the local structures of the PC are neglected. From convolutional neural networks (CNNs), we know that leveraging the local structure is a key aspect of the success of CNNs, in which local features are extracted from small neighborhoods,  grouped into larger units, and processed to produce higher level features. This is the well-known ``hierarchical feature extraction'', deeply used in 2D computer vision and processing tasks.   In PCs,  neighboring points form a meaningful subset that captures key semantic information about the 3D geometry; hence they should retain even more information than the 2D counterpart. Because of this intuition, PointNet++~\cite{qi2017pointnet++} introduced a hierarchical architecture for PC processing, capturing features at increasingly larger scales along a multi-resolution hierarchy. 
This concept is illustrated in Figure~\ref{fig:teaser}, where a PC input is processed at different scales (middle part of the figure) to extract hierarchical features (right side of the figure) at different levels.  At the lower level (``Local'' in the figure), each point neighborhood covers a small and densely populated region, extracting fine geometric structures.   In contrast, at the higher levels (``Global" in the figure), the network captures coarser structures from larger neighborhoods. 
\begin{figure*}[t!]
\centering
\includegraphics[width=1.0\textwidth]{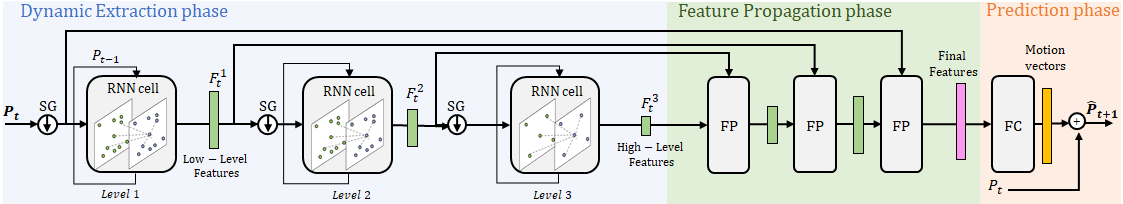}
\caption{\textit{Classic} hierarchical architecture of Graph-RNN with three levels for the prediction task. 
}
\label{fig:big_shechamtic}
 \vspace*{-5mm}
\end{figure*}
Given the increasing importance of \textit{dynamic} PC sequences in a 
wide spectrum of applications from automation to virtual reality,
several works have extended the PointNet++ network by introducing spatio-temporal neighborhoods in order to learn temporal features.
Most of these works adopt hierarchical architectures~\cite{fan2022point,fan2022pstnet,Graph-RNN,FlowNET3D,PointRNN},
which can be considered the de-facto approach for  dynamic point cloud processing today.

The common intuition is that such hierarchical architecture allows learning more descriptive features, pushing researchers to develop even more hierarchical (and possibly complex) models.  However, 
 \textit{why}  such models work and \textit{which} features do they learn in the framework of PC processing is still not understood and usually overlooked in the literature. 
Initial understanding has been provided for the PointNet model~\cite{qi2017pointnet,zhang2019explaining}.
However, such efforts are limited to the original PointNet, which does not have a hierarchical learning architecture and it processes static PCs only, leaving a gap in the understanding of current PC processing models.    
For example,  which motion or flow is learned at the different stages of the hierarchical architecture is unknown. Also which key components of the networks (multi-scaling, stacking of deep nets, etc.) lead to the success of the model remain overlooked. 
Bringing more clarity to such hierarchical learning models when applied to dynamic PC sequences is the goal of this work, which provides a clear understanding of the hierarchical spatial-temporal features extraction in dynamic PCs. 

Specifically, we perform an experimental study using state-of-the-art prediction architecture via hierarchical Recurrent Neural Networks (RNN) proposed in Point-RNN~\cite{PointRNN} and improved in Graph-RNN~\cite{Graph-RNN}.
To this end, we design different variations of the hierarchical architecture and perform an ablation study across architectures using a dataset of human body motion~\cite{Mixamo}, with the clear intent of disentangling the multi-scale effect from the deep learning one. Then, we explain the learned features for each tested architecture by visualizing the body motion vector.  This strategy allows us to demonstrate how learning features at multiple scales is equivalent to extracting local and global temporal correlations from data.  This confirms what has been studied in CNN architectures~\cite{zeiler2014visualizing}. 
However, unlike CNNs, in which only the last layer features are processed to infer the final task, we show that all the learned features are important to properly capture complex movements (and perform downstream tasks).  In fact, we show that either when the hierarchical learning is not considered or when is considered but only the last level is taken into account, we cannot translate anymore the local and global features to local and global movements and the final task is not properly inferred.  
While some of these insights might be intuitive, to the best of our knowledge this is the first work validating those intuitions in the dynamic PC setting. Moreover, we believe that understanding the effect of the different components of the network on the features and the prediction accuracy   can  inform future research directions in how to learn more representative hierarchical features as well as how to take advantage of their combination towards the development of simpler and more accurate methods.

\section{Background - Hierarchical Architecture }
Without loss of generality, we focus on the architecture depicted in  Figure~\ref{fig:big_shechamtic} for the task of PC prediction~\cite{PointRNN,Graph-RNN}. 
The architecture, which extracts the spatio-temporal features  via RNN cells, is composed by two phases: 
%
%

\begin{enumerate}
    \item Dynamic Extraction phase (DE):   the network processes the input  PC frame $P_t$ and extracts its hierarchical features $F_t$;
    \item Feature Propagation phase (FP):  combines the learned features  to reconstruct the predicted point cloud $\hat{P}_{t+1}$, which is the  PC at the next time step.
    \end{enumerate}
We now describe both phases in more details.

\subsubsection{Dynamic Extraction phase (DE)}
The DE phase takes a PC $P_t$ (pre-processed or raw) as input and extracts the PC dynamic behavior.
The DE phase consists of multiple \textit{stacked} RNN cells, for a total of $L$ levels. Before being processed by the RNN cell, the PC is downsampled by a \textit{Sampling and Grouping} (SG) module, as described in~\cite{qi2017pointnet++}.  
Hence, a progressive subsampled PC is fed into subsequent RNN cells. 
%
In each RNN, each point of the cloud is processed jointly with its spatio-temporal neighborhood, defined as its $k$-closest neighbors in space and time. It is worth noting that due to the subsequent sampling (hence a sparser PC  at later levels/RNN cells) leads to a neighborhood with larger distances between points.
As a result, the first level learns local features $F^1_t$ from small scale neighborhoods whereas the last level learns global features $F^L_t$ 
observing large scale neighborhoods.
On top of such multi-scale effect, the stacked RNNs make the model deeper imposing that features learned at each level $l$ are the input to the next level $l+1$. 
In natural language processing (NLP) the advantage of staking RNN with respect to the vanilla RNN model has been empirically proved~\cite{pascanu2014construct}.
In this work, we aim at understanding if the gain from the stacked effect also remains in PC processing.

\subsubsection{Feature Propagation phase (FP)}
Once DE phase has learned the features from all the levels ($F_t^{1},F_t^{2},... ,F_t^{L}$), the FP phase combines them
into a single final feature ($F_t^{Final}$).
The combination is done by hierarchically propagating the features from the higher levels to the lower levels through several Feature Propagation (FP)~\cite{qi2017pointnet++} modules. 
This is done by first interpolating the sub-sampled features from the higher level to the same number of points as the lower level, followed by a concatenation of the interpolated features with lower-level features.
The concatenation is then processed by a point-based network and a ReLU.
The process is repeated in a hierarchical manner until the features from all the levels have been combined into a final feature.
As last step, after the FP phase, the final features are converted into motion vectors using a fully connected layer. The calculated motion vectors are then added to the original PC to predict the PC at the next time step.

It is worth noting that while the prediction phase is specific for the final application (i.e., prediction)
the DE phase, complemented by the FP phase, is the core of a hierarchical PC processing strategy 
which is common to other downstream tasks, such as classification~\cite{fan2022point} and  segmentation~\cite{ fan2022pstnet}.



\section{Experimental Study}

In this section, we present the different architectures along with  the dataset and simulation settings  used in our study. 

\subsection{Dataset}
Similarly to~\cite{Graph-RNN}, we use the human  \textit{Mixamo}~\cite{Mixamo} dataset, consisting of sequences of human bodies while performing various dynamic activities such as dancing or playing sports.  The main motivations behind this selection are given in the following: compared to real 3D scenes acquired by LiDAR  or mmWave sensors, such synthetic dataset does not suffer from  acquisition  noise or  quantization distortion. This makes the dataset easy to visualize and hence to  comprehend, while also isolating the problem of learning features from noisy corrections and other aspects that may arise in a more noisy dataset.  
Additionally,  the synthetic dataset allow us to generate very different movements (from quite simply like walking to highly complicating such as breakdance steps) and study the learning of the highly descriptive features in such cases.



\subsection{Experimental Architectures}\label{SubSec_Architectures}
\begin{figure}[t]
		\centering
		\subfigure[\textit{Shallow} hierarchical architecture]{
			\centering
			\includegraphics[width=0.47\textwidth]{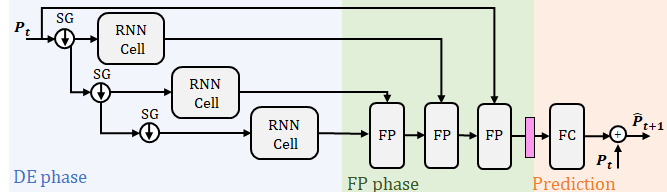}         \label{fig:aq_2}}
		\subfigure[\textit{Single-scale} hierarchical architecture]{
			\centering
			\includegraphics[width=0.47\textwidth]{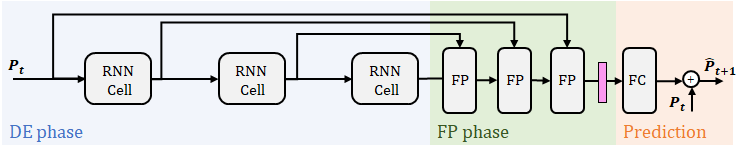}         \label{fig:aq_3}}
		\subfigure[\textit{Without-combination} hierarchical architecture]{
			\centering
			\includegraphics[width=0.47\textwidth]{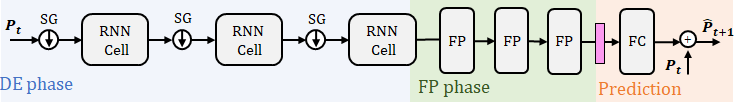}         \label{fig:aq_4}}
		\caption{Experimental architectures.
		\label{fig:schemes}
		}
	\end{figure}

To better understand and  explain the hierarchical features learning,  different architectures have been implemented, trained and compared  to the \textit{Classic} hierarchical architecture depicted in Figure~\ref{fig:big_shechamtic}. The key idea is to isolate the multi-scale resolution, from the ``deepness" of the architecture, disentangling in such a way    the different aspects of the PC processing.  The common aspects of all implemented solutions is the high level architecture depicted in Section~II: a DE phase (with $L=3$ and  downsampling factor in SG of four), a FP phase and a final prediction step. On the other side, the models differentiate  in the RNN processing (stacked or parallel), as well as in the sampling modules. Specifically, we implemented the following architectures:
\begin{itemize}


     \item \textit{Shallow} hierarchical architecture -- Figure~\ref{fig:schemes}\,a): as in  \textit{Classic} architecture, the PC is sub-sampled in a stacked fashion, leading to a very sparse PC at level $L=3$. Unlike the  \textit{Classic} architecture, the RNN cells are in parallel (instead of stacked), leading to a Shallow network (local features are not grouped and processed at higher levels). This network highlights the effect of multi-scaling (sampling) instead of deep hierarchical learning. 
     
     \item \textit{Single-scale} architecture -- Figure~\ref{fig:schemes}\,b): Same architecture as in \textit{Classic} hierarchical model but without the  downsampling modules at each level. Hence, all the levels process the same number points and learn features at the same scale.

     \item \textit{Without-combination} architecture -- Figure~\ref{fig:schemes}\,c): this architecture recall the classic deep neural network in which  local features are extracted from small neighborhoods (first RNN cell) and processed at higher (subsequent) level. Only the last level feature is then used for the final reconstruction.  Note that in the  \textit{Classic} architecture all features learned at different level are used for  predicting the final motion vectors. 
\end{itemize}

\subsection{Results Evaluation and Visualization}
The above  models, as well as the \textit{Classic} one, are trained using as loss function a combination of the Chamfer distance (CD)~\cite{fan2017point} and earth's moving distance (EMD)~\cite{fan2017point}  to measure the distance between predicted $\hat{P}_{t+1}$ and the target PC  $P_{t+1}$, as explained in~\cite{Graph-RNN}. Those metrics are also used for the  evaluation of PC prediction in the  experimental results discussion. It is worth noting that the CD distance tends to flatten all scores toward zero values. This is because in the PC the majority of the points are perfectly predicted (all points with no motion or little motion) and most of the errors (high CD scores) are focused in the high motion area. This is the area of strongest interest in PC prediction tasks as we are interested in understanding if the neural network is able to capture such movements. Therefore, we also consider the CD Top 5\% metric, which looks at the CD metric of the 5\% points with the worst prediction (\ie points with the farthest distance to their closest point). 

Besides the aforementioned metrics, we also  visualize  the features learned at each level as motion vectors. This is a key aspect to better understand the hierarchical learning process, the goal of this paper. As shown in Figure~\ref{fig:big_shechamtic}, the motion vectors are obtained by combining features from multiple levels in the FP phase. The final features are then processed by a last fully connected layer and converted into motion vectors. We are interested in visualizing the actual contribution to the final motion vectors from each level. We do this by seeing the motion vectors  as the combination of motion vectors produced at each level \ie $M_t =\sum_{l}^L M_t^l$, 
with  $M_t$ being the predicted motion vectors and $M_t^l$  the motion vectors produced by level $l$.  Such level contribution $M_t^l$ is visualized by keeping the features from level $l$ and setting to zero the remaining ones at the input of the FP phase. We then 
replicate the FP and prediction phase operation with the already trained weights and obtain the motion vector generated by level $l$. 

\section{Explaining hierarchical features}
We now present and discuss the results obtained from our study of hierarchical features, which is the core contribution of this work. 
We aim at explaining \textit{what} features are  learned
and \textit{how} these are learned in the \textit{Classic} hierarchical architecture. Then,  we investigate the role that  the combination of hierarchical features has in the dynamic PC processing (by comparing the \textit{Classic} with those presented above).


\subsection{Hierarchical Learning}
\begin{figure}[t!]
\centering
\includegraphics[width=0.495\textwidth]{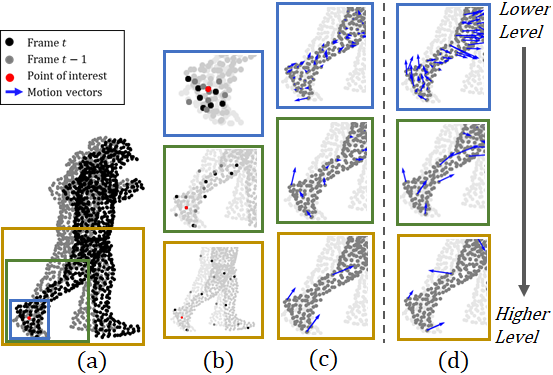}
 \caption{(a) Input dynamic PC; (b) Multi-scale neighborhood at different levels;   motion vectors  learned at  level  $l$ with (c) \textit{Classic}  and  (d) \textit{Shallow}  networks.}
\label{fig:local_global}
\end{figure}

To study hierarchical learning in dynamic PC processing, we study the features learned from the  \textit{Classic} architecture in Figure~\ref{fig:big_shechamtic}. We visualize those features giving as input PC   a running person shown in Figure~\ref{fig:local_global}. 
We investigate how each level of the architecture learns features for a given point of interest, such as a point in the foot (red point in the Figure~\ref{fig:local_global}). To do so, we first show in Figure~\ref{fig:local_global}\,b) the multi-scale neighborhood of the selected point in the foot at each level.
Due to the subsequent sampling in the network, the lower level learns features only by looking at points in a small area around the point of interest (top blue square in the figure). 
On the contrary, the higher level learns features by considering a sparser set of points in a large area (bottom gold square in the figure). 
In Figure~\ref{fig:local_global}\,c), we visualize the  motion vectors $M_t^l$ at each level of the \textit{Classic} hierarchical architecture for   given time frame $t$.
It can be observed that the lowest level captures small and diverse motions; on the contrary, at the highest level, a single main motion vector is learned for almost all the points in the foot. 
Similarly to what happens in hierarchical CNNs, we observe that the highest level of the network learns global motions (for example the forward motion of the runner), while lower levels of the architecture learn local movements of the action (small refinement motion that defines the local movement of the foot -- which can also go up-backward when the foot goes up). While intuitive, to the best of our knowledge, this visualization is the first one in the literature confirming that the intuition behind hierarchical features in CNNs can be extended to dynamic PCs architectures and dynamic flow: \textbf{hierarchical features can be explained as local and global motions.}

To carry out our deeper understanding of hierarchical learning, now we want   to answer the following questions:
\textit{How does the hierarchical architecture learn global and local motion?} (Sections\,\ref{subsection:cascading} and \ref{subsection:multiscale}); \textit{How does the hierarchical PC architecture differ from   CNN?} (Section\,\ref{subsection:featcomb}).

\subsection{Is Stacked Effect the key?}
\label{subsection:cascading}
To understand better how global and local motions are learned in the hierarchical architecture, we first investigate the effect of the stacked component on the learned features. To do so, we compare the \textit{Classic} architecture (in which local features are aggregated and processed at a higher level) to the \textit{Shallow} one, 
which learns features without features aggregation. Note that in both architectures there is the multi-scaling effect, with the higher lever observing a much sparser PC (hence processing a more global neighborhood).
Looking at the motion vectors $M_t^l$ learned in  the \textit{Shallow} (Figure\,\ref{fig:local_global}\,d)) architecture, we can note that  1) motion vectors are all at the same magnitude across levels; 2)  at higher levels  the vectors show highly contrasting movements. This implies that in the \textit{Shallow} architecture the features lose the global and location motion interpretation. Specifically, we do not have the global motion (runner moving forward) being captured by the higher levels. Similarly, lower levels do not capture only small and refining local motions. This is motivated by the fact that
in the \textit{Shallow} architecture all three levels contribute equally to the output motion.
This proves that the features interpretation is not guaranteed by the  multi-scale component in the architecture, while it is learned by the stacked aspects of local features being aggregated and processed to learn global motion.  

\begin{figure}[t!]
\centering
\includegraphics[width=0.48\textwidth]{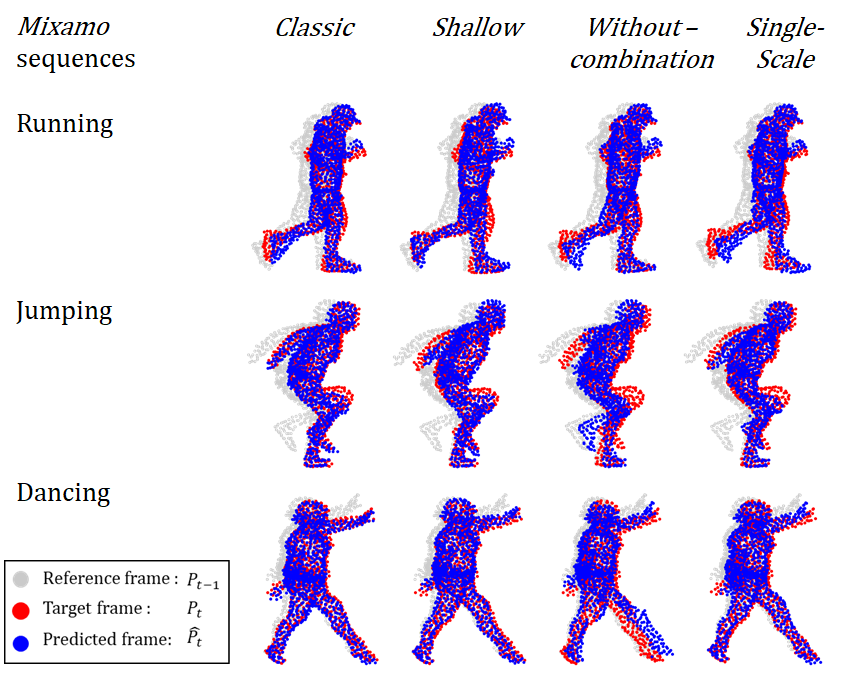}\caption{Predicted PCs in the different architecture for three sequences of the \textit{Mixamo} dataset.}
\label{fig:visual_result}
 \vspace*{-5mm}
\end{figure}

Interestingly, we have also observed that despite the missing interpretation of global and local motion, the \textit{Shallow} can capture the overall movement of the PC and perform a decent prediction. This is observed in
Table~\ref{tab:architectures}, which shows   the prediction error for the  different architectures presented in Section\,\ref{SubSec_Architectures}. We also add the prediction error for the ``Copy-Last-Input" case, in which the prediction is simply the copy of the previous input. This shows that all the methods investigated in this work perform substantially better than ``Copy-Last-Input", hence they are all able to capture some aspects of the body movements. We also visualise in  Figure~\ref{fig:visual_result} the prediction for three particular sequences: \textit{Running, Jumping, and Dancing}. We can observe that the Dancing PC is very well predicted from both \textit{Shallow} and \textit{Classic} architectures. In Jumping and Running sequences, there is a mild gain from the \textit{Classic} one (see for example the arm in Jumping) but the \textit{Shallow} architecture still learns well the overall movement. 
In short, while the stacked aspect has a big impact on the interpretation of the learned feature, the performance gap is not as substantial.

%

\subsection{Is Multi-Scale Effect the key?}
\label{subsection:multiscale}
We are now interested in understanding how much the multi-scale effect is important for capturing the PC motion. 
We compare the \textit{Classic} architecture to the \textit{Single-scale}  architecture (\ie without
multi-scale effect) and we show that the multi-scale is essential to achieve a good performance (hence prediction). This is confirmed both numerically and visually.  In Table\,\ref{tab:architectures},  the \textit{Classic} architecture has a better overall prediction accuracy and  
from  Figure~\ref{fig:visual_result}, it is clear that \textit{Single-scale}  architecture fails in capturing some local movements such as the foot/leg in ``Running", the knee in ``Jumping" and the hands in ``Dancing".  
This is motivated by the following: in complex movement, as in the ``Running", in which the foot performs a forward translation and rotation motion simultaneously (Figure~\ref{fig:local_global}\,c), the joint-motion can be captured only by looking at points from different scales, the forward movement required a large-scale neighborhood for example. Only by observing the foot neighborhoods at multiple scales both the global forward motion and local rotation motion are captured. This ability to capture complex motion patterns as local and global motions results in more accurate predictions.




\subsection{Is This Like Any Other Deep Learning?} 
\label{subsection:featcomb}
We are now interested in answering the second question: \textit{How does the hierarchical PC architecture differ from   CNN?} Both architectures use a hierarchical approach to learn global and local features.
However, in CNNs only the last layer features are usually processed to infer the final task, while in the \emph{Classic} architecture  all  features from all the levels are combined in an FP propagation phase. We then compare the \emph{Classic} architecture to the \textit{Without-combination} one, which does not combine the features from different levels. From Figure~\ref{fig:visual_result}, we can already observe that the \textit{Without-combination} architecture fails in predicting key local movements: knee in ``Jumping" and leg in ``Dancing". Even if the network  can  learn hierarchical features, it is not able to merge the local motions in the final reconstruction, in which only the higher level motion is used for the prediction.  As a result, the network loses the ability to predict complex motions, resulting in an inferior performance (validated  in Table\,\ref{tab:architectures}).

\begin{table}[t!]
\centering
\caption{Prediction error in all the \textit{Mixamo} test sequences.}
\resizebox{0.4\textwidth}{!}{%
\begin{tabular}{|c|ccc|}
\hline
\multirow{2}{*}{Experimental Architectures} & \multirow{2}{*}{CD} & \multirow{2}{*}{EMD} & \multirow{2}{*}{\begin{tabular}[c]{@{}c@{}}CD\\ Top 5\%\end{tabular}} \\
                                            &                     &                      &                                                                       \\ \hline
\textit{Classic}                            & \textbf{0.00262}    & \textbf{59.6}        & \textbf{0.1412}                                                       \\ \hline
\textit{Shallow}                            & 0.00268             & 62.2                 & 0.1459                                                                \\ \hline
\textit{Single-scale}                       & 0.00331             & 65.0                 & 0.1609                                                                \\ \hline
\textit{Without-combination}                & 0.00314             & 67.3                 & 0.1752                                                                \\ \hline
\textit{Copy-Last-Input}                    & 0.01056             & 123.4                & 0.2691                                                                \\ \hline
\end{tabular}%
}

\label{tab:architectures}
 \vspace*{-5mm}
\end{table}




\section{Conclusion}
By performing an experimental study on state-of-art architectures for PC prediction, this   paper explains hierarchical features in the context of dynamic PC processing. Specifically, we have shown 1) the interpretation of low- and high-level features as local and global motions;  2) the importance of different components of the networks in current  state-of-the-art models to achieve a better PC predictions. We believe that such insights can open the door to new designs of more efficient and accurate networks for future PC processing tasks such as learning-based PC compression algorithms.




\bibliographystyle{IEEEtran}

\bibliography{main}

\end{document}